\documentclass[]{aipproc}
\layoutstyle{8x11double}

\begin{document}
\title[Short Gamma-ray Bursts in Close Neutron Star Binaries]{A Model for Short Gamma-Ray Bursts: Heated Neutron Stars in Close Binary Systems}

\author{Jay D. Salmonson}{
	address={Lawrence Livermore National Laboratory, P.O. Box 808,
	Livermore, CA 94551}, email={salmonson@llnl.gov},
	homepage={http://members.home.net/jdsalmonson} }

\author{James R. Wilson}{
	address={Lawrence Livermore National Laboratory, P.O. Box 808, Livermore, CA 94551},
	email={jimwilson@llnl.gov}
}

\copyrightyear  {2001}

\begin{abstract}
In this paper we present a model for the short (< second) population
of gamma-ray bursts (GRBs).  In this model heated neutron stars in a
close binary system near their last stable orbit emit neutrinos at
large luminosities ($\sim 10^{53}$ ergs/sec).  A fraction of
these neutrinos will annihilate to form an $e^+e^-$ pair plasma wind
which will, in turn, expand and recombine to photons which make the
gamma-ray burst.  We study neutrino annihilation and show that a
substantial fraction ($\sim 1/2$) of energy deposited comes from
inter-star neutrinos, where each member of the neutrino pair
originates from each neutron star.  Thus, in addition to the
annihilation of neutrinos blowing off of a single star, we have a new
source of baryon free energy that is deposited between the stars.  To
model the $e^+e^-$ pair plasma wind between stars, we do
three-dimensional relativistic numerical hydrodynamic calculations.

Preliminary results are also presented of new, fully general
relativistic calculations of gravitationally attracting stars falling
from infinity with no angular momentum.  These simulations exhibit a
compression effect.
\end{abstract}

\date{\today}

\maketitle

\section{Introduction}

In \citet{swm01} a model was presented for the production of a GRB in
a close neutron star binary system, near its last stable orbit.  In
that model the stars undergo compression due to non-linear general
relativistic effects.  Vortices and shocks within the stars will
convert this compressional energy into thermal energy which will be
radiated from the stars in neutrinos.  These neutrinos emerging from
the neutron stars will partially recombine via $\nu\overline{\nu}
\rightarrow e^+e^-$, an effect that is substantially augmented (up to
30 times) by bending of neutrino paths by strong gravitational fields
\citep{sw99}.  Thus an $e^+e^-$ pair plasma fireball emerges from the
neutron stars and expands relativistically.  A key parameter studied
in \citet{swm01} was the entropy per baryon, $s$, of the plasma,
representing the amount of baryons entrained in the $e^+e^-$ fireball.
Ita was found through 1D relativistic hydrodynamic simulations that if
$s \sim 10^8$, then prompt gamma-ray emission would result from the
eventual recombination of $e^+e^- \rightarrow \gamma\gamma$.  If the
entropy is as low as $s \sim 10^6$, effectively all of the energy
would be transferred as kinetic energy to the baryons, thus resulting
in gamma-ray emission from an external shock as the relativistic
baryons sweep into the interstellar medium.  Each scenario was studied
in detail in \citet{swm01}.

In the current work we employ three enhancements.  The first is from
recent refined simulations of the compression by Wilson \& Mathews
which show timescales $\sim 1$ second, thus this model is best suited
to describe the short class of GRBs (< second).  Second, we use new
calculations by \citet{sw01} of the $\nu\overline{\nu} \rightarrow
e^+e^-$ between the neutrons stars where each neutrino of the
annihilation pair originates from each star.  This effect is found to
be of about the same importance as that of the previously considered
annihilation from individual netron stars \citep{sw99}.  Thus we have
a new source of baryon-free $e^+e^-$ plasma between the neutron stars.
Third, we implement fully relativistic 3D hydrodynamics to model the
plasma expansion in the complex, rotating inter-star environment.

\section{The Model}

\begin{figure}
\includegraphics[height=.35\textheight]{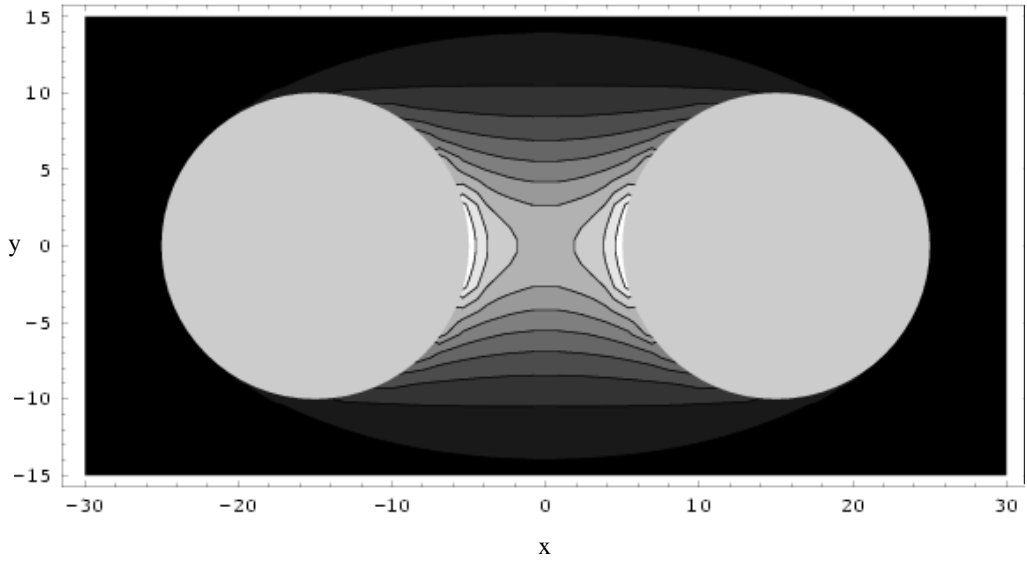}
\caption{A contour plot of interstar neutrino annihilation energy
depostion rates on a slice through a neutron star binary system
\citep{sw01}, with stellar radii 10 km and separation 30 km. Lighter
colors correspond to higher levels of energy deposition where black
includes zero.
\label{contourplot}}
\end{figure}

In this model we estimate that 10\% of the binding energy of a neutron
star ($\sim 10^{53}$ ergs) is converted to thermal energy via
compression, vortices and shocks.  This $10^{52}$ ergs of energy is
released as a monotonically increasing luminosity of neutrinos over a
timescale of order $\sim 1/10$ second as found by calculations of
J.~R.~Wilson \& G.~J.~Mathews.  This neutrino luminosity $\sim
10^{53}$ ergs sec$^{-1}$ annihilates into an $e^+e^-$ pair plasma.
Annihilation of high neutrino luminosities in the strong gravitational
field of the neutron stars can have high efficiencies; near unity
\citep{swm01}.  About half of the pair plasma energy is deposited
uniformly around the neutron stars due to single star neutrino
annihilation \citep{sw99} (Figure \ref{contourplot}) and the other
half is deposited between the stars due to interstar neutrino
annihilation \citep{sw01}.

\begin{figure}
\includegraphics[height=.3\textheight]{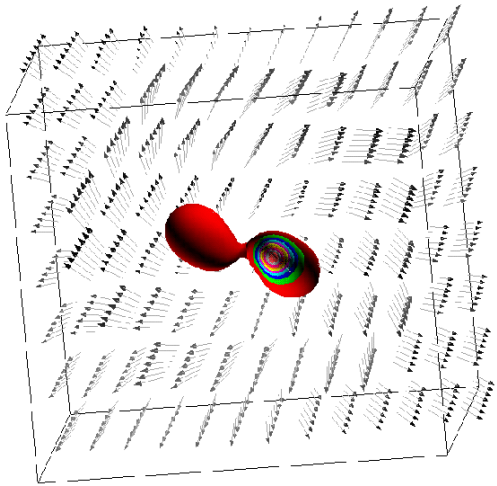}  
\caption{Three dimensional relativistic simulation of two 10 km radius
neutron star separated by 30 km emitting $10^{53}$ ergs/sec of energy
in $e^+e^-$ pair plasma and about 1 \% equivalent mass in baryons.
The contour map, with right star cutaway, is of baryon density. The
vector field is the 3-velocity of expanding plasma.  This problem
settles down to a static flow after about one orbit with period $\sim
1/300$ second.
\label{plot3d}}
\end{figure}

This plasma deposition morphology then becomes input for the 3D
relativistic hydrodynamic code, which calculates the expansion of the
plasma (Figure \ref{plot3d}).  These simulations show a plasma of very
high entropy expanding out along the plane of symmetry between the
neutron stars.  In the regions around the stars lower entropy plasma
is formed because a baryon wind is blown from the stars.

Thus this model predicts a variety of bursts.  Viewed along the axis
of rotation, a prompt quasi-thermal burst of duration $\sim 1/10$
second will result from the annihilation of fireball pairs.  Because
of the dearth of baryons left over to sweep into the interstellar
medium, we do not predict the existence of an afterglow.  This agrees
with preliminary searches of the data archives for short-burst
afterglows, which appear to be missing \citep{g+01}.

Viewed far from the axis of rotation a very different burst results.
The lower entropy means that there will not be a prompt burst from
pair annihilation in the fireball.  However, there will be a baryon
wind sweeping into the interstellar medium.  Thus we expect a burst
that decays into an afterglow as a power-law.  This behavior will be
made chaotic and complex by the rapid rotation of the binary system.

\section{The Compression Effect in Strongly Gravitating Systems}

In \citet{wmm96} it was reported that neutron star binaries near their
last stable orbit undergo a compression due to non-linear general
relativistic (GR) effects.  This effect could be strong enough to
crush the stars to black holes and perhaps release binding energy as
thermal neutrinos in the process.  Since that report, this effect has
remained controversial.

\begin{figure}
\includegraphics[height=.25\textheight]{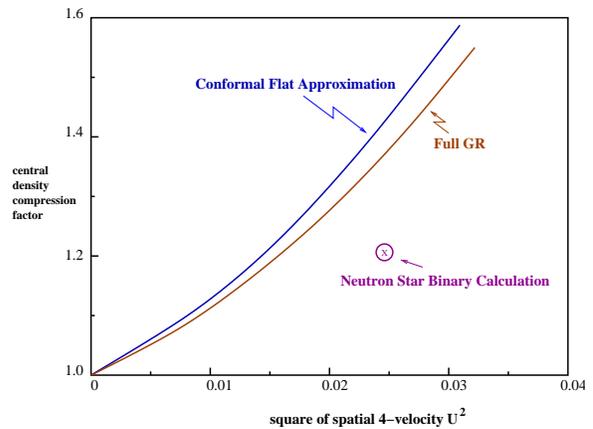}
\caption{A schematic diagram of the increase in central density as a
function of the spatial 4-velocity of two close neutron stars observed
in three different calculations.  The point marked with an 'X' shows
the magnitude of the compression calculated by \citet{mw00} in neutron
star binary calculations at their last stable orbit.  The curves show
two calculations for two neutron stars falling toward eachother from
infinity, with no angular momentum.  One curve is done in full general
relativity (GR), made possible by the axisymmetry of this problem, and
the other with the conformal flat approximation (CFA) used in the
neutron star binary calculations \citep{mw00}.  One can see that all
the CFA does a good job of reproducing the solution of the exact, full
GR calculations.  Also, we see that compression is observed for both
rotating and linear systems.
\label{centraldensity}}
\end{figure}

Wilson recently has done a similar calculation of two stars
gravitationally falling together without angular momentum.  The
calculations were done both in full GR and using the ``conformal flat
approximation'' (CFA), often cited by critics to be the spurious
source of the compression effect.  Preliminary results, schematically
shown in Figure \ref{centraldensity}, demonstrate a compression effect
for both calculations.  Thus, not only is the CFA not the source of
the compression, but perhaps the compression effect, here demonstrated
in a non-rotating system, is more general than previously thought.

This work was performed under the auspices of the U.S. Department of
Energy by University of California Lawrence Livermore National
Laboratory under contract W-7405-ENG-48.

\bibliographystyle{arlobib}
\def\apjl{ApJ}
\def\apj{ApJ}
\def\nat{Nature}
\def\physrep{Phys.~Rep.}
\def\prd{Phys.~Rev.~D}

\end{document}